\documentclass[aps,pra,amsmath,amssymb,showpacs,twocolumn,longbibliography]{revtex4-1}
\usepackage[T1]{fontenc}
\usepackage{amssymb,amsmath,bm}
\usepackage{amsfonts}
\usepackage{graphicx}
\usepackage{amsthm}
\usepackage{color}
\usepackage{mathbbol}
\usepackage{hyperref}
\usepackage{cleveref}
\usepackage{epstopdf}
\usepackage{mathrsfs}
\usepackage{placeins}
\usepackage{tikz}
\usepackage{verbatim}
\usepackage{mathtools}
\usepackage{preamble_revPRA}
\usepackage{paralist}

\begin{document}

\title{Many-body quantum metrology with scalar bosons in a single potential well} 

\author{Julien M.\,E. Fra\"isse, Jae-Gyun Baak, and Uwe R. Fischer}
\affiliation{Seoul National University, Department of Physics and Astronomy, Center for Theoretical Physics, Seoul 08826, Korea}

\begin{abstract} 
We theoretically investigate the possibility of performing high precision estimation of an externally imposed acceleration using scalar bosons in a single-well trap. We work at the level of a two-mode truncation, valid for weak to intermediate two-body interaction couplings. 
 The splitting process into two modes is in our model entirely caused by the interaction 
 between the constituent bosons and is hence neither due to an externally imposed double-well potential nor due to  populating a spinor degree of freedom.
The precision enhancement gained by using various initial quantum states using a two-mode bosonic system is well established.  Here we therefore instead focus on the effect of the intrinsic dynamics on the precision, where, in a single well, the Hamiltonian assumes a form different from that of the typical double-well case. We demonstrate how interactions can significantly increase the quantum Fisher information maximized over initial states  as well as the quantum Fisher information for a fragmented or a coherent state, the two many-body states that can commonly represent the ground state of our system.
\end{abstract}

%\pacs{03.67.-a,03.67.Lx}
\maketitle
%\begin{twocolumn}

%%%%%%%%%%%%%%%%%%%%%%%%%%%%%%%%%%%%%%%%%%%%%%%%%%%
%%%%%%%%%%%%%%%%%%%%%%%%%%%%%%%%%%%%%%%%%%%%%%%%%%%
\section{Introduction}\label{sec:intro}
%%%%%%%%%%%%%%%%%%%%%%%%%%%%%%%%%%%%%%%%%%%%%%%%%%%
%%%%%%%%%%%%%%%%%%%%%%%%%%%%%%%%%%%%%%%%%%%%%%%%%%%

%Due to the probabilistic nature of 
{Within quantum theory, it has been possible to apply --- and extend --- results from classical statistics to what is now known as quantum parameter estimation theory, whose aim is to investigate questions related to the estimation of a deterministic parameter --- say $\lambda$ --- using a quantum system \cite{giovannetti_quantum_2004}. }The main paradigm is, having $\nu$ copies of a $\lambda$-dependent quantum state $\rho_\lambda$ at our disposal, to find what the best possible precision is, according to quantum mechanics, for the estimation of $\lambda$. The quantum \crr/ bound answers this question by relating the variance of any unbiased estimator $\lambda_\mm{est}$ to the quantum Fisher information (QFI) \cite{helstrom_minimum_1967,helstrom_quantum_1976,braunstein_statistical_1994}: \begin{equation}
 \Vcl{\lambda_\mm{est}} \geq \frac{1}{\nu \qfinue_\lambda}\,.
\end{equation}  
Crucially, this bound can always be saturated in the limit of large $\nu$, therefore making the QFI $\qfinue_\lambda$ a proper  figure of merit for the precision which can in principle be attained.

In general, a dynamical protocol for quantum metrology is four-step: The preparation of an initial state, the imprinting of the parameter during the evolution, the measurement of the  final state, and eventually --- after repeating these three  steps $\nu$ times --- the estimation of the parameter from the measurement results \cite{giovannetti_quantum_2006}.  An important class of quantum protocols are based on interferometry, and many interferometers can be modeled as a Mach-Zehnder interferometer (MZI), which is a two-mode interferometer. In the optical case --- the original one --- the two modes correspond to the two arms of the interferometer. Light is sent to one or both input ports, goes through a first beam splitter, accumulates a phase in one arm, and goes through a second beam splitter. Eventually a measurement at the output of the arms is made to collect the data from which the parameter is inferred. Using a classical state at the input the QFI is at most equal to $N t^2$, the shot noise limit, while by using properly designed states one can reach a QFI equal to $N^2t^2$, known as the Heisenberg limit \cite{holland_interferometric_1993,zwierz_general_2010,zwierz_ultimate_2012}.

{In this optical scenario, the parameter $\lambda$ is imprinted by a phase-shift Hamiltonian, i.e.~a Hamiltonian $\lambda \hat{G}$ in which the parameter $\lambda$ is an overall multiplicative factor, and where $\hat{G}$ is the generator of the transformation. For a long time the study of quantum metrology has been focusing on such Hamiltonians. Recently, a lot of attention has been devoted in quantum metrology to Hamiltonians which, having a more complex dependency on the parameter, are not phase-shift Hamiltonians \cite{boixo_generalized_2007,boixo_quantum_2008,boixo_quantum_2009,tacla_nonlinear_2010,tikhonenkov_optimal_2010,grond_mach_zehnder_2011,javanainen_optimal_2012,javanainen_ground_2014,de_pasquale_quantum_2013,pang_quantum_2014,pang_optimal_2017,fraisse_enhancing_2017,skotiniotis_quantum_2015}.
For example, while $\lambda(\hj_z + \chi \hj_x)$ and $\lambda \hj_z^2$ are phase-shift Hamiltonians with respect to $\lambda$, $\cos(\lambda) \hj_x + \sin(\lambda) \hj_y$ and $ \lambda \hj_z + \hj_x$ are not phase-shift Hamiltonians. Notice, in particular, that nonlinear Hamiltonians \emph{can} be phase-shift Hamiltonians (see, 
e.g., \cite{luis_nonlinear_2004,luis_quantum_2007,luis_precision_2010}).} 
It turns out that when considering complex Hamiltonians, not all of them allow the same precision in quantum metrology. Also the question of the time behavior (which is trivial for a phase-shift Hamiltonian) becomes more intricate here. Such studies show the importance of the dynamics, and demonstrate that it is insufficient to only discuss the influence of the initial state to assess the 
precision attainable.

Beyond the optical case, quantum metrology can be based on atomic systems as well \cite{keith_interferometer_1991,riehle_optical_1991,kasevich_atomic_1991,wineland_spin_1992,wineland_squeezed_1994,bollinger_optimal_1996}, 
and in particular on cold quantum gases \cite{gross_spin_2012,lee_nonlinear_2012,huang_quantum_2013,pezze_quantum_2016,toth_quantum_2014}. For example, using two-mode bosonic atomic systems it is also possible to design an MZI, where  the beam splitting process depends on the particular system. These systems can be represented by an assembly of qubits in a symmetric state, the internal states of spinor gases, or cold gases in a double-well trap, see for a selection of the extensive literature \cite{huelga_improvement_1997,dunningham_interferometry_2002,gerry_resolution_2003,jaaskelainen_limits_2004,shin_atom_2004,pezze_sub_shot_noise_2005,pezze_phase_2006,dalton_two_mode_2007,dalton_two_2012,gietka_atom_2014,pezze_phase_sensitivity_2015,skotiniotis_quantum_2015}.
  We notice that metrology with cold atoms has already been implemented  experimentally, e.g., in Refs.~\cite{gross_nonlinear_2012,berrada_integrated_2013,
 van_frank_interferometry_2014,sewell_ultrasensitive_2014}. Formally, all these two-mode systems can be mapped to a spin system using the SU(2) representation. Then, a priori, results obtained for a specific system can be directly applied to another system, in particular regarding the enhancement offered by a given initial state. Nevertheless this is true only as long as the dynamics governing the systems are identical. 
Hence we focus in our study below on the role of the dynamics for the estimation precision.

In this paper we investigate how scalar bosons in a single-well trap can be used for estimating an external force (an acceleration) applied to the bosons (the estimation of the intrinsic coupling parameters of the Hamiltonian has  been considered in \cite{volkoff_optimal_2016}), and in particular to assess the effect of interaction. We work in a regime of weak to intermediate 
interaction couplings, for which the two-mode truncation approximately applies. Contrary to the double-well case where the splitting into two modes corresponds to a spatial splitting, in our scenario the splitting is purely coming from interactions. It is also distinct from a spinor gas in a single trap where the ``splitting'' is performed internally, and the particles occupy two hyperfine states (with the usual approximation that they share the same spatial mode). For a double well, the natural description by two modes is established by a sufficiently large barrier, and the dynamics is represented by the competition of the single-particle tunneling between these modes and the interactions in each well. The single-well geometry imparts definite parity to the modes, thus the system has {\em pair-tunneling} dominating over the negligible single-particle tunneling.  In addition, there occurs a term involving the product of occupation numbers in the modes; both of the latter two terms are exponentially small in the double-well geometry.
Therefore the dynamics of our system in particular as regards the crucial interplay between the mode occupations, is fundamentally different from the one obtained in the double-well scenario. %The effect of nonlinear dynamics on precision has previously been considered in the optical %and the cold atom cases in 
%Refs.~\cite{luis_nonlinear_2004,luis_quantum_2007,luis_precision_2010,boixo_quantum_2008,boixo_quantum_2009,tacla_nonlinear_2010,tikhonenkov_optimal_2010,grond_mach_zehnder_2011,javanainen_optimal_2012,javanainen_ground_2014}.

Our paper is organized as follows. In section \ref{sec:phys_model} we introduce the physcial system, its Hamiltonian, discuss the main differences between single and double well geometries and present the different scenarios for metrology. In section \ref{sec:cQFI} we first introduce the tools to study quantum metrology in presence of a nontrivial dynamics, in particular the maximal QFI and an upper bound to it. We then use these tools to analyze the effect of the interaction on the precision of the estimation of the external force, and show how, in the most realistic (within our model) regime of parameter (small accelerations), interactions are needed to reach a high QFI. In section \ref{sec:ground},  to explore more realistic scenarios, we look at the performance of a fragmented and a coherent states on a protocol resembling an MZI. The analysis shows again how interaction couplings can enhance the estimation precision of the acceleration.

%%%%%%%%%%%%%%%%%%%%%%%%%%%%%%%%%%%%%%%%%%%%%%%%%%%
%%%%%%%%%%%%%%%%%%%%%%%%%%%%%%%%%%%%%%%%%%%%%%%%%%%
\section{Scalar bosons in a single well}\label{sec:phys_model}
%%%%%%%%%%%%%%%%%%%%%%%%%%%%%%%%%%%%%%%%%%%%%%%%%%%
%%%%%%%%%%%%%%%%%%%%%%%%%%%%%%%%%%%%%%%%%%%%%%%%%%%

\subsection{System Hamiltonian and ground states}\label{H} 
We consider a harmonically trapped ultracold 1D bosonic gas with short-ranged two-body interactions of coupling strength $g_{\rm 1D}$. Introducing a rescaled coupling constant
\begin{equation}
g=N g_\mm{1D},
\end{equation}  the Hamiltonian can be written 
\begin{equation}
\hat{H}_\mm{sys}=\frac{1}{2}\sum_{\alpha=1}^{N}\bigg[-\frac{\partial^{2}}{\partial x_{\alpha}^{2}}+x_{\alpha}^{2}\bigg]
	+\frac{g}{2N} \sum_{\alpha,\beta}\delta(x_{\alpha}-x_{\beta})	\,,
\end{equation}
setting $\hbar$ and atomic mass $m$ both to unity, and 
where the coordinates of atoms are denoted by $x_{\alpha}$. 
For notational convenience, we rescale energies by  
$l^{-2}$ and lengths by $l$, which represents the harmonic oscillator length. 
All quantities in what follows are thus dimensionless (given in units of powers of $l$).  
Introducing the bosonic field operator $\hat{\psi}(x)$ that satisfies the commutation relations $\com{\hat{\psi}(x)}{\hat{\psi}^\dagger(x')}=\delta(x-x')$ and $\com{\hat{\psi}(x)}{\hat{\psi}(x')}=\com{\hat{\psi}^\dagger(x)}{\hat{\psi}^\dagger(x')}=0$ gives the field-quantized form of $\hat{H}_\mm{sys}$:
\begin{multline}
\hat{H}_\mm{sys}=\frac{1}{2}\int dx\,\hat{\psi}^{\dagger}(x)\Big[-\frac{\partial^{2}}{\partial x^{2}}+ x^{2}\Big]\hat{\psi}(x)\\ +\frac{g}{2N}\int dx\,\hat{\psi}^{\dagger}(x)\hat{\psi}^{\dagger}(x)\hat{\psi}(x)\hat{\psi}(x)\,.\label{eq:2ndquantham} 
\end{multline} 
We work at the level of a two-mode truncation. Writing the two-mode truncated field operator  as $\hat{\psi}(x) = \psi_{0}(x)\hat{a}_{0}+\psi_{1}(x)\hat{a}_{1}$ with annihilation operators $\hat{a}_{0}$ and $\hat{a}_{1}$, we obtain  \cite{bader_fragmented_2009}
\begin{multline}
\hat{H}_\mm{sys}=\sum_{i=0}^{1}\epsilon_{i}\hat{a}^{\dagger}_{i}\hat{a}_{i}
+\frac{ A_{1}}{2}\hat{a}^{\dagger}_{0}\hat{a}^{\dagger}_{0}\hat{a}_{0}\hat{a}_{0}+\frac{ A_{2}}{2}\hat{a}^{\dagger}_{1}\hat{a}^{\dagger}_{1}\hat{a}_{1}\hat{a}_{1} \\ +\left[\frac{ A_{3}}{2  }\hat{a}^{\dagger}_{0}\hat{a}^{\dagger}_{0}\hat{a}_{1}\hat{a}_{1}+\textrm{h.c.}\right]
+\frac{ A_{4}}{2}\hat{a}^{\dagger}_{0}\hat{a}_{0}\hat{a}^{\dagger}_{1}\hat{a}_{1}\,,\label{eq:mainham}
\end{multline}
where the single-particle energies are given by 
$\epsilon_{i}  = \frac{1}{2}\int dx\,\psi_{i}^{\ast}(x)\big[-\frac{\partial^{2}}{\partial x^{2}}+ x^2\big]\psi_{i}(x)$, and the  interaction couplings read $A_{1} =  V_{0000}$, $A_{2}= V_{1111}$, $ A_{3}= V_{0011} $, and $ A_{4}=V_{0101}+V_{1010}+V_{1001}+V_{0110}$, where  $V_{ijkl} = (g/N)\int dx\,\psi_{i}^{\ast}(x)\psi_{j}^{\ast}(x)\psi_{k}(x)\psi_{l}(x)$.
$A_{1}$ (resp.~$A_{2}$) is the interaction energy of atoms in $\psi_{0}(x)$ [resp.~$\psi_{1}(x)$],  $A_{3}$ is the amplitude for pair tunneling, and $A_{4}$ is the intermode density-density interaction coupling. We stress again that both $A_3$ and $A_4$ are exponentially smaller than $A_1$ and $A_2$ in a double well. 
The orbitals $\psi_{0}(x)$ and $\psi_{1}(x)$ are assumed to be real for the simplicity of our discussion, which in turn renders all $A_i$ also real. Due to the symmetry of the trap potential  (e.g. of the above specified harmonic type) $\psi_{0}(x)$ and $\psi_{1}(x)$ have a spatial parity: $\psi_{0}(x)$ is even while $\psi_{1}(x)$ is  odd. %Then, terms of the form 
%$\hat{a}_{0}^{\dagger}\hat{a}_{1}$ (particle exchange) or $\hat{a}_{0}^{\dagger}\hat{a}_{0}^{\dagger}\hat{a}_{0}\hat{a}_{1}$ (occupation number-weighed particle exchange) do 
Then, terms corresponding to particle exchange 
(e.g.~$\propto\,\hat{a}_{0}^{\dagger}\hat{a}_{1}$) or occupation number-weighed particle exchange 
(e.g.~$\propto\,\hat{a}_{0}^{\dagger}\hat{a}_{0}^{\dagger}\hat{a}_{0}\hat{a}_{1}$) do 
not contribute to the Hamiltonian. 

For scalar bosons in a single-well trap, the justification of  the two-mode truncation is not as straightforward as for bosons in a double well, 
where it is the spatial separation that 
at least suggests the use of a two-mode truncation, by simply using the respective ground-state orbitals of left and right well \cite{milburn_quantum_1997}.
A more refined variant of the two-mode approximation uses properly chosen effective modes, see \cite{Sowinski}
for a detailed analysis and further references .However, one should realize that the two-mode truncation has its limits of validity 
also in a double well, cf.~the self-consistent analysis of \cite{Sakmann}.  
In a harmonic trap, for very weak interaction --- the single-condensate regime --- the system is well described using a single mode (that is, by single-orbital mean-field theory), while for very strong interaction --- the Tonks-Girardeau regime --- the system is fermionized and its description requires a very large  
number of modes \cite{alon_pathway_2005}. In this context, the two-mode truncation is the first step to take into account the effect of interaction beyond the mean-field regime. To obtain an estimate regarding the regime of validity of our model we use some existing results concerning our model.  In \cite{dunjko_bosons_2001} the authors characterize the single-condensate and the Tonks-Girardeau regimes. A Lieb-Liniger-type parameter $\gamma= g_{\rm 1D}/n$ may be introduced to identify and discriminate both regimes. Here, $n$ represents the central density.
% in the Thomas--Fermi limit. 
With our choice of units the Lieb--Liniger type parameter approximately reads $\gamma \simeq 1.5\times g_{\rm 1D}^{4/3}N^{-2/3}$. In the single-condensate regime the parameter $\gamma \ll 1$  while in the Tonks-Girardeau regime $\gamma\gg 1$. Here we consider the two-mode truncation to be valid in the intermediate regime, that is for values of $\gamma$ of the order (or below) one,  so that $g_{\rm  1D}^{4/3} \lesssim N^{2/3}$, which translates as $g \lesssim N^{3/2}$. 
We emphasize that this simple analysis does not thoroughly justify the two-mode truncation of the field operator expansion.
A more complete analysis determining in particular the energy level occupation statistics must involve self-consistent many-body calculations cf., e.g.,  the multiconfigurational Hartree approach utilized in \cite{alon_pathway_2005,Marchukov}.

\subsection{SU(2) representation and parameter regimes}\label{sec:su2}

There exists an important connection between a spin system and a two-mode bosonic system. Namely, the latter is formally equivalent to a spin of size $J = N/2$.
In the Schwinger representation, we define the SU(2) operators as
$\hj_x=(\hat{a}_{0}^{\dagger}\hat{a}_{1}+\hat{a}_{1}^{\dagger}\hat{a}_{0})/2$,
$\hj_y=(\hat{a}_{0}^{\dagger}\hat{a}_{1}-\hat{a}_{1}^{\dagger}\hat{a}_{0})/2 \ii$,
$\hj_z=(\hat{a}_{0}^{\dagger}\hat{a}_{0}-\hat{a}_{1}^{\dagger}\hat{a}_{1})/2$,
and $\hj_0=(\hat{a}_{0}^{\dagger}\hat{a}_{0}+\hat{a}_{1}^{\dagger}\hat{a}_{1})/2=\hat{N}/2$.
Then, the Hamiltonian of the bosons in a single trap within the two-mode truncation is recast as
\begin{equation}\label{eq:lmg_ham1}
\hat{H}_\mm{sys}=- \de \hj_z + g\left[\frac{N-1}{2N} \dA \hj_z +  \frac{\eta}{N}(\hj_x^2 + \xi \hj_y^2) \right]\,.
\end{equation}
In terms of the parameters introduced  in Eq.~\eqref{eq:mainham}, we have
\begin{align}
\de &= \epsilon_1-\epsilon_0 \,, \label{de}\\
\dA &=\tilde{A}_1-\tilde{A}_2 \,, \label{dA}\\
\eta &  =   \frac{\tilde{A}_4+2 \tilde{A}_3-\sigma_{\tilde{A}}}{2} \,,\label{eta} \\
\xi &  =  \frac{\sigma_{\tilde{A}}+2 \tilde{A}_3-\tilde{A}_4}{\sigma_{\tilde{A}}-(2\tilde{A}_3+\tilde{A}_4)}\,, \label{xi}
\end{align} 
with $\sigma_{\tilde{A}}=\tilde{A}_1+\tilde{A}_2$, 
$\tilde{A}_{i}=A_{i}/g_{\mm{1D}}$, and where $\de$ corresponds to the energy difference between the single-particle modes while $\eta$, $\xi$, and $\dA$ characterize the interaction terms. We see that the interaction Hamiltonian has two contributions: a nonlinear term and a renormalization of the single-particle term. To make this last point more apparent we write the Hamiltonian as
\begin{equation}\label{eq:lmg_ham}
\hat{H}_\mm{sys}=q \hj_z + \eta \frac{g}{N}(\hj_x^2 + \xi \hj_y^2)\,.
\end{equation}
with 
\begin{equation}
q   = g\frac{(N-1)}{N}\frac{\dA}{2}-\de  \label{q}
\end{equation}
the renormalized energy difference.

 The assumed reality of the mode functions implies that $\tilde{A}_{i}\geq 0$, $\tilde{A}_{4}=4\tilde{A}_{3}$, and $\sigma_{\tilde{A}}\geq 2\tilde{A}_{3}$.  Further, these relations imply that
\begin{enumerate}
\item[(i)]  $\eta$ and $\xi$ have opposite signs,
\begin{equation}
\mm{sgn}(\eta)=-\mm{sgn}(\xi)\,,
\end{equation} 
\item[(ii)]  $\xi$ is negative or larger than unity,
\begin{equation}
\xi  \leq 0 \quad \text{or} \quad \xi  \geq 1\,,
\end{equation}
\end{enumerate}
where $\xi=0$ is achieved for $\psi_0(x)=\psi_1(x)$ and $\xi=1$ is achieved for $\psi_0(x)\psi_1(x)=0$. No restrictions apply a priori to $\dA$. Note that in a double well, to exponential accuracy in the overlap of the two modes
in the left/right wells, we always have $\xi = 1$ and $\eta <0$.

While the value of $g$ is restricted by physical considerations (see introduction) on the regime of validity of our model ($g\lesssim N^{3/2}$), the situation is less transparent for the parameters $\eta$, $\xi$, and $\dA$. They purely depend on the spatial integral of the mode functions. As we need to be specific in order to perform numerical calculations, we choose the values of these parameters in agreement with the values obtained when using the ground and first excited  state mode functions of a single particle in a harmonic oscillator potential. In particular this choice of mode function implies that $\tilde{A}_2=\frac34 \tilde{A}_1$, $\tilde{A}_3=\frac12 \tilde{A}_1$, and $\tilde{A}_4=2\tilde{A}_1$ \cite{bader_fragmented_2009}. With $\tilde{A}_1=1$ we obtain $\eta=0.625$, $\dA=0.25$, and $\xi=-0.6$, which are the values we use in the rest of the paper if not specified otherwise. As observed above, these $\eta$ and $\xi$ have signs 
{\em opposite} to those of a double well and $|\xi|\neq 1$.
Because we later on use the fact that the ground state of the system is fragmented, we restrict our analysis to a small number of particles ($N=50$), because the two-mode fragmentation degree approximately decreases  as $N^{-1/2}$,  
all other parameters fixed \cite{condensate_fischer_2015}.

\subsection{Two-mode metrology with a Bose gas}

Our metrological protocol is designed to estimate the value of an externally imposed acceleration which generates the Stark-type potential $\chi \sum_{\alpha=1}^{N} x_{\alpha}$ in the Hamiltonian, where $\chi$ is the force. Its field-quantized form is 
\begin{equation}
\hat H_{\rm acc} = \chi \kappa(\hat{a}_{0}^{\dagger}\hat{a}_{1}+\hat{a}_{1}^{\dagger}\hat{a}_{0}), 
\label{acc}
\end{equation}
with $\kappa  = \bra{0}\hat{x}\ket{1}=\bra{1}\hat{x}\ket{0}=\int dx\, \psi_{0}^{\ast}(x)x\psi_{1}(x)$. Also, the SU(2) representation of  $\hat H_{\rm acc}$  is 
\begin{equation}
\hat H_{\rm acc}(\lambda)=\lambda \hj_x\,,
\end{equation}
 with $\lambda  = 2\chi\kappa$. In the following $\lambda$ will be the parameter we want to estimate and is referred to as the acceleration. To estimate the value of the acceleration, one starts by preparing an initial state $\ket{\psi_0}$, lets it evolve unitarily with Hamiltonian $\hat{H}_\mm{sys}+\hat{H}_\mm{acc}(\lambda)$, and eventually performs a measurement on the final state in order to infer the value of $\lambda$ from the measurement results. To quantify the precision of the protocol, we base our analysis on the QFI and therefore do not consider the question of the measurement itself.
 
It is important at this point to discuss the differences between our system and the more frequently studied double-well system. The typical Hamiltonian studied in the double-well case is \cite{grond_mach_zehnder_2011,javanainen_optimal_2012,gietka_atom_2014,javanainen_ground_2014}

 \begin{equation}\label{eq:ham_dw}
\hat{H}_\mm{dw}=\de \hj_z +\Omega \hj_x + u \hj_z^2\,,
\end{equation}
 where $\de$ is also the level difference, $\Omega$ is the tunneling rate, and $u$ is the interaction strength. In the double-well Hamiltonian, the terms $A_3$ and $A_4$ present in the single-well Hamiltonian [see equation \eqref{eq:2ndquantham}] are negligible and $A_1=A_2$ which means that no renormalization of the single-particle term occurs ($\dA=0$). On the other hand, in the single-well Hamiltonian, there is no tunneling $\Omega$ due the definite parity of the modes.   In the absence of interaction, the double-well Hamiltonian has exactly the same form as  our Hamiltonian in presence of the acceleration. Formally the metrological analysis would go through identically. The crucial differences appear in the presence of interaction. While in the double-well system interaction produces only a nonlinear term of fixed form, in our system we obtain a renormalization of the single-particle term plus a nonlinear term, whose form depends on the mode functions. Notice also that many authors \cite{javanainen_optimal_2012,grond_mach_zehnder_2011,gietka_atom_2014,javanainen_ground_2014}
 have been focusing on the estimation of $\de$ in the Hamiltonian \eqref{eq:ham_dw}. Then the generator $\hj_z$ commutes with the interaction $\hj_z^2$. In our system the situation is more complex as in general the interaction term ($\hj_x^2+\xi \hj_y^2$) does not commute with the generator ($\hj_x$).
 
 In terms of protocol, different situations have been considered within the double-well scenario.  In what we call the \emph{ideal} protocol (or ideal interferometer), which requires a full control of the parameter $\de$ and $\Omega$, no interactions are present, and the beam splitting is performed by applying only the $\hj_x$ operator, while the phase accumulation is performed by applying only the $\hj_z$ operator (for the estimation of $\de$) \cite{pezze_phase_2006}. This is formally the exact equivalent of the optical MZI. One can also consider a more complex dynamics, still assuming no interactions. Then the relevant Hamiltonian for phase accumulation is $\de \hj_z +\Omega \hj_x$ \cite{gietka_atom_2014}. In general, $\hj_x$ has a detrimental effect in terms of precision. Finally one can consider the full dynamics using the Hamiltonian \eqref{eq:ham_dw}  to assess the effect of interactions \cite{javanainen_optimal_2012}.

Can these three different scenarios be reproduced within our model for the estimation of an acceleration?  The ideal scenario is not actually realistic  since we cannot set $\de=0$ in a single potential well.
It is neither realistic to consider the regime $\de \ll \lambda$ which would effectively realize an ideal interferometer.  It is more appropriate to consider the regime $\lambda \lesssim \de$ when we take  into account that the value of $\lambda$ plays also a role when discussing the regime of validity of the two-mode truncation: For large $\lambda$ comparative to $\de$, we may expect a third mode to get populated.

The noninteracting scenario is a priori accessible. There one would prepare an initial state in the presence of interaction and later on turn off the interaction using a Feshbach resonance. 
However, the interacting scenario is the most realistic, as it is the natural configuration of the system. Furthermore,  even when using a Feshbach resonance to reduce interactions, it is not guaranteed that there is no residual interaction remaining between the bosons.

%\col{We now turn our attention to the channel QFI, to illustrate the predominant importance of the form of the  Hamiltonian and of the presence of interactions in our setup.}

%%%%%%%%%%%%%%%%%%%%%%%%%%%%%%%%%%%%%%%%%%%%%%%%%%%%
%%%%%%%%%%%%%%%%%%%%%%%%%%%%%%%%%%%%%%%%%%%%%%%%%%%
\section{Channel QFI}\label{sec:cQFI}
%%%%%%%%%%%%%%%%%%%%%%%%%%%%%%%%%%%%%%%%%%%%%%%%%%%
%%%%%%%%%%%%%%%%%%%%%%%%%%%%%%%%%%%%%%%%%%%%%%%%%%%

We start  the analysis by using the \emph{channel} QFI (cQFI), which is nothing else than the QFI optimized over the set of initial states --- that is why we also call it \emph{maximal} QFI. The interest of this figure of merit is to keep the focus on the dynamics, on the Hamiltonian, as the explicit state dependency of QFI is cleared out.  Indeed, the cQFI is  an upper bound to the QFI: If the cQFI is low, then, independent from the initial state, the QFI will be low. When considering the cQFI the main problem for achieving high quantum metrological precision lies in the dynamics and not in the initial state. Consequently, for an increase of the QFI the dynamics must be modified. This makes the cQFI a figure of merit that suits our purpose particularly well, as we want to investigate how the dynamics affects the quantum enhancement in the precision of the estimation of the force.

\subsection{Hamiltonian parameter estimation}

Consider a generic Hamiltonian $H(\lambda)$ depending on a parameter $\lambda$ that we want to estimate. We define the cQFI as the QFI maximized over all possible initial states  \cite{fujiwara_fibre_2008}:
\begin{equation}
\cqfinue_\lambda =\max_{\ket{\psi_0}} \qfinue_\lambda (\e{-\ii t \hat{H}(\lambda)}\ket{\psi_0})\,,
\end{equation}
where $\qfinue_\lambda (\e{-\ii t \hat{H}(\lambda)}\ket{\psi_0})$ is the QFI for the state $\e{-\ii t \hat{H}(\lambda)}\ket{\psi_0}$. The cQFI can be expressed conveniently introducing the dynamical generator $\hat{\msc{H}}$ \cite{giovannetti_quantum_2006,pang_quantum_2014} defined as 
\begin{equation}
\hat{\msc{H}} = \ii \hat{U}^\dagger \partial_\lambda\hat{U}\,,
\end{equation}
where $ \hat{U}$ is the evolution operator $\hat{U}=\e{-\ii t \hat{H}(\lambda)}$. The dynamical generator has also been introduced in the context of double-well metrology \cite{javanainen_optimal_2012}, and a closed form of it depending on the spectral decomposition of the Hamiltonian has been found \cite{pang_quantum_2014}, allowing to infer some properties especially in terms of time scaling (see for a detailed discussion below). Using the dynamical generator we have
\begin{equation}
\cqfinue_\lambda = \pnorm{\hat{\msc{H}}}{\rm SN}^2\,, \label{cQFI_SN}
\end{equation}
with the semi-norm $\pnorm{\bullet}{\rm SN}$ defined as the difference between the maximal and the minimal eigenvalues of the operator \cite{boixo_generalized_2007,fraisse_enhancing_2017}.

It is important to notice that the cQFI is upper bounded as follow: 
\begin{equation}
\cqfinue_\lambda \leq t^2\pnorm{\partial_\lambda \hat{H}(\lambda)}{\rm SN}^2\,.
\end{equation}
Such an upper bound can be generalized to time-dependent Hamiltonians \cite{pang_optimal_2017}. The exact conditions for the saturation of the bound have been given in \cite{fraisse_enhancing_2017}. In particular, these conditions are fulfilled when the Hamiltonian commutes with its first derivative, which is the case for a phase-shift Hamiltonian.

\subsection{Noninteracting system}

\begin{figure}
\centering\includegraphics[scale=1]{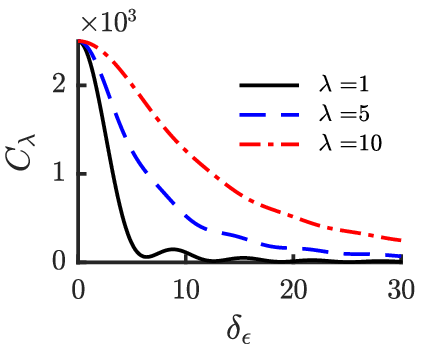}\hfill\includegraphics[scale=1]{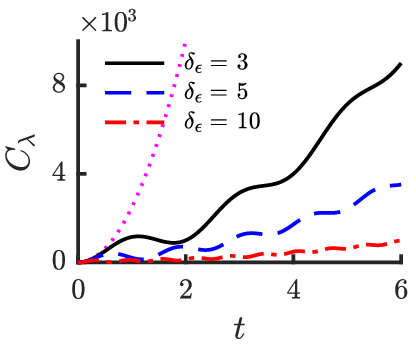}
\caption{Maximal QFI (cQFI) [equation \eqref{eq:cQFI_noint}] for a noninteracting system (the left plot has $t=1$ and the right plot has $\lambda=1$). The dynamics is $\lambda \hj_x - \de \hj_z $ and the parameter to be estimated is $\lambda$. The plots illustrate the detrimental effect of the presence of the $\de \hj_z$  term during the phase accumulation (for large $\de$ the cQFI becomes very small).  }\label{fig:cQFI_noint}
\end{figure}

In our model, the relevant Hamiltonian during the phase accumulation stage is $\hat{H}(\lambda)=\hat{H}_\mm{sys} +\hat{H}_\mm{acc}(\lambda)$, and its derivative is simply $\hj_x$. Therefore the upper bound of the cQFI takes the form $\cqfinue_\lambda \leq t^2N^2$, which is the Heisenberg limit. This upper bound is nothing else than the cQFI for an ideal interferometer. Therefore \emph{with respect to the  cQFI} the ideal interferometer has an optimal dynamics.

But as we discussed in the above, the ideal protocol cannot be realized using scalar bosons in a single-well trap. There is always the single-particle term remaining, and in the absence of interaction the Hamiltonian is $\hat{H}(\lambda)\vert_{g=0} = \lambda \hj_x -\de \hj_z$, with $\de$ the energy difference. This configuration was thoroughly studied in the context of the double well \cite{gietka_atom_2014}. As it is possible to diagonalize the Hamiltonian, we can explicitly write down the dynamical generator $\hat{\msc{H}}\vert_{g=0}$ and calculate its semi-norm to obtain the maximal QFI: 
\begin{equation}\label{eq:cQFI_noint}
\cqfinue_\lambda = N^2 \Big\lbrace\frac{t^2 \lambda^2}{\lambda^2+\de^2}+\left(\frac{2\de}{\lambda^2+\de^2}\right)^2 \sin^2\left(\frac{t}{2}\sqrt{\lambda^2+\de^2})\right) \Big\rbrace \,.
\end{equation} 

We can check in equation \eqref{eq:cQFI_noint} that by taking $\de=0$ we retrieve the result for the ideal protocol (the Heisenberg limit). For small $\de$ (keeping in mind that for too low values of $\de$ the two-mode approximation in our model becomes unrealistic) the cQFI is equal to $N^2(t^2-\mu \de^2)+\mc{O}(\de^4)$ with $\mu\geq 0$, demonstrating the detrimental effect of the $\hj_z$ operator. This is illustrated by the left-hand plot in the figure \ref{fig:cQFI_noint}, where we see how near the origin the cQFI decreases for increasing $\de$. For larger  $\de$ values the cQFI is not a monotonous function of $\de$ anymore (see the black straight line corresponding to $\lambda = 1$). Such a nonmonotonic behavior --- meaning that far from the limiting point, we do not know if it is better to increase or reduce the detrimental term in the Hamiltonian --- has  been observed  previously in \cite{de_pasquale_quantum_2013,fraisse_enhancing_2017}.  %\col{We also notice that for $\lambda=0$ the cQFI is not equal to zero but to $4 N^2\sin^2(t\de/2)/\de^2$.}\rmk{Actually useless statement. If anybody is interested in this question it takes him one second to set $\lambda=0$ and to look at the resulting expression}  

Finally an interesting point as regards the cQFI is its time scaling. The time scaling of the cQFI is well understood in the context of Hamiltonian parameter estimation. It is composed by a quadratic term and an oscillating one \cite{pang_quantum_2014}. 
The quadratic term finds its origin in the parameter dependence of the \emph{eigenvalues} of the Hamiltonian while the oscillating  term has its origin in the parameter dependence of the \emph{eigenvectors}. In particular a phase-shift Hamiltonian has only a parameter dependence on its eigenvalues, and its QFI scales as $t^2$. In the nonideal interferometer, the presence of the $\hj_z$ term in the Hamiltonian results in an oscillating time behavior of the cQFI, as illustrated in the right-hand plot of the figure \ref{fig:cQFI_noint}.

Overall, we have thus found that the presence of the  $\hj_z$ term in the dynamics has a detrimental effect on the precision. As we commented above, within the two-mode truncation one   should keep $\de$ larger than $\lambda$, and equation \eqref{eq:cQFI_noint} demonstrates that under this condition the cQFI is highly suppressed. The cQFI being an upper bound to the QFI, it follows that in this parameter regime it is not possible to fully exploit the quantum enhancement, {\em irrespective} of the choice of the initial state. To restore quantum enhancement, it is necessary to modify the dynamics, for example by including the two-body interaction between the bosons.
 
\subsection{Effect of interactions on the channel QFI}

\begin{figure}
\centering\includegraphics[scale=0.95]{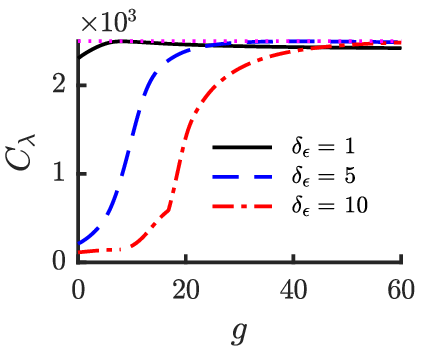}\hfill\includegraphics[scale=1]{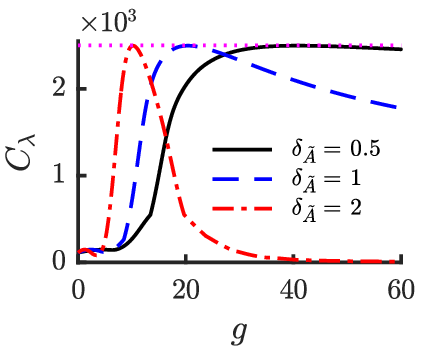}
\centering\includegraphics[scale=0.95]{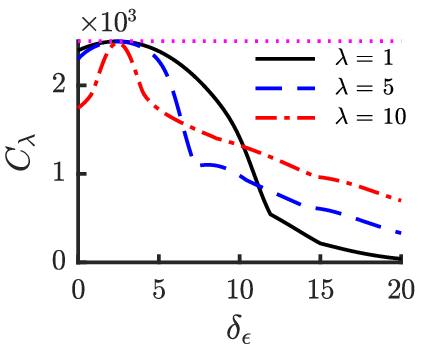}\hfill\includegraphics[scale=1]{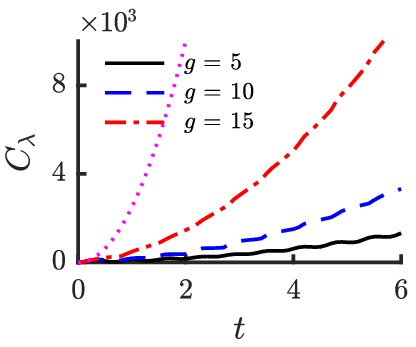}

\caption{Maximal QFI (cQFI) in the presence of interaction. The magenta dotted line displays the upper bound $N^2t^2$ (obtained by using an ideal interferometer). We see in the upper left plot ($\lambda=1$ and $t=1$) how a large enough interaction strength helps to achieve a high cQFI, almost reaching its upper bound. On the upper right plot ($\lambda=1$, $t=1$, and $\de=10$) we illustrate how for larger values of $\dA$ only near $q =0$ the cQFI saturates its upper bound. On the bottom plots we investigate how the behavior as a function of $\de$  (left plot, $t=1$ and $g=20$) and $t$ (right plot, $\lambda=1$ and $\de = 10$)  is modified by the presence of interactions (compare with figure \ref{fig:cQFI_noint}).  }\label{fig:cQFI_various}
\end{figure}

We will now study the effect of the interactions on the channel QFI. Such a study has previously also been realized in a double well for an  estimation of $\de$ \cite{javanainen_optimal_2012}.
Notice that then the interaction term commutes with the generator ($\hj_z^2$ and $\hj_z$ respectively). In the limit of large interaction the authors of  \cite{javanainen_optimal_2012} 
showed that the cQFI saturates its upper bound. In our case the situation is more complex as the interaction introduces a renormalization of the single-particle energy as well as a nonlinear term which does not commute with the generator. As we mentioned in the section \ref{sec:su2}, we use the values obtained for harmonic oscillator mode functions for our parameters $\eta$, $\xi$, and $\dA$, and keep $g$ as a free parameter.

In the upper left-hand plot on figure \ref{fig:cQFI_various} we represented the cQFI as a function of $g$ for different values of $\de$. For $\de=1$ the cQFI is already high without interactions, grows somewhat for low values of $g$ and then decreases slightly. The most interesting results are found with higher $\de$ values which, importantly, corresponds to the most realistic regime --- in terms of the acceleration --- for the validity of the two-mode truncation. Then we already know that in the absence of interaction the cQFI is very low. Here we see that interaction helps to tremendously increase the cQFI, and to almost saturate the upper bound (dotted magenta line). This result is similar to the one obtained for the double well, although our situation is a priori less favorable (as the nonlinear term does not commute with the generator).

In our model, the ratio $\eta/\dA$ is important as it controls [assuming that 
$|\xi|\simeq \mc{O}(1)$] the relative weight between single-particle and nonlinear term. With our choice of mode function we have $\eta/\dA=2.5$, so the nonlinear term dominates.  To get an idea of the effect of a deviation from this ratio we plotted in the top right of the figure \ref{fig:cQFI_various} the cQFI as a function of $g$ again for different values of $\dA$. For large values of $\dA$, the cQFI reaches its upper bound for lower values of $g$, but then decreases (and the larger is $\dA$, the faster it decreases). The point were the cQFI reaches its maximum corresponds to the point were the single-particle term vanishes, $q=0$, which is achieved roughly at $g$ equal to $2\de/\dA$. As we see from this formula, the higher $\dA$, the lower is the optimal $g$. %\colb{But at the same time, when $\dA$ dominates over $\eta$ and  $g$ is large, the Hamiltonian is effectively dominated by the $\hj_z$ operator attached to $\dA$. And we know that $\hj_z$ has a detrimental effect. This analysis is in agreement with the observation of the steep decrease of the cQFI with moderate $g$ when $\dA$ dominates over $\eta$ (see for example the curve corresponding to $\dA=2$)}. 

Finally, in the bottom row of the figure \ref{fig:cQFI_various} we plotted the cQFI as a function of $\de$ (left) and $t$ (right) to directly compare with the noninteracting case (see figure \ref{fig:cQFI_noint}). Regarding $\de$, we again see how, near the point $q=0$, the cQFI almost saturates its upper bound (here at $\de \simeq 2.5$). We also observe in which way the dynamics can produce some nontrivial result, as for low values of $\de$, increasing $\lambda$ does not necessarily lead to a higher cQFI (e.g.~at $\de=5$). Regarding the $t$ dependence, we witness how the presence of interaction reduces the oscillating behavior, and increases the quadratic part.

%%%%%%%%%%%%%%%%%%%%%%%%%%%%%%%%%%%%%%%%%%%%%%%%%%%%
%%%%%%%%%%%%%%%%%%%%%%%%%%%%%%%%%%%%%%%%%%%%%%%%%%%
\section{Metrology with ground states }\label{sec:ground}
%%%%%%%%%%%%%%%%%%%%%%%%%%%%%%%%%%%%%%%%%%%%%%%%%%%
%%%%%%%%%%%%%%%%%%%%%%%%%%%%%%%%%%%%%%%%%%%%%%%%%%%

We now turn our attention  to a more realistic (that is, practically realizable) scenario, and look at the performance of the ground state of the system for the estimation of the acceleration $\lambda$. In the spin language, the ground state $\ket{\psi_\mm{g}(\theta)}$ of the scalar bosons  in a single trap is a superposition, for sufficiently large $N$, of two spin coherent states:
\begin{equation}\label{eq:ground_state}
\ket{ \psi_\mm{g}(\theta)}  = \frac{1}{\sqrt{2}}\Big(\ket{\theta,\frac{\pi}{2}} + \ii \ket{\theta,\frac{3\pi}{2}}\Big)\,,
\end{equation}
with $\ket{\theta,\phi}$ a spin coherent state where $\theta$ is the azimuthal angle and $\phi$ the polar angle  \cite{bader_fragmented_2009,fischer_robustness_2013,kang_revealing_2014}. In terms of bosonic algebra we have $\ket{\theta,\phi}  = \frac{1}{\sqrt{N!}}\big[\cos(\theta/2)\hat{a}_{0}^{\dagger}+\e{\ii \phi}\sin(\theta/2)\hat{a}_{1}^{\dagger}\big]^{N} \ket{0,0}$  with $\ket{0,0}$ the vacuum state.  Superposition of spin coherent states has already been studied in the context of quantum metrology (see for example \cite{huang_quantum_2015}), but not for the particular dynamics of interacting scalar bosons in a single trap we consider here. 

Physically, the angle $\theta$ parametrizes the degree of fragmentation according to $\dof=2\sin^{2}(\theta/2)$ for $0\leq\theta\leq\pi/2$ or $\dof=2-2\sin^{2}(\theta/2)$ for $\pi/2\leq\theta\leq\pi$. The degree of fragmentation $\dof$ is defined as $\dof=1-\vert\lambda_{0}-\lambda_{1}\vert/N$ where $\lambda_0$ and $\lambda_1$ are the two eigenvalues of the single-particle density matrix \cite{mueller_fragmentation_2006}. In \cite{bader_fragmented_2009,fischer_robustness_2013} the exact form of $\theta$ depending on the parameters of the Hamiltonian  \eqref{eq:lmg_ham1} was calculated.
However, the values of $\theta$ obtained only using the Hamiltonian parameters and the 
harmonic oscillator ground and first excited states as orbitals are unrealistically high (cf.~the fully self-consistent results of \cite{condensate_fischer_2015}).  
Therefore we assume an arbitrary value of $\theta$ that corresponds to a low degree of fragmentation, namely $\theta=0.5$ ($\dof\simeq 0.12$). We also consider a coherent state obtained  by setting $\theta=0$ (and $\dof=0$ as well) as a representative of an almost vanishing degree of fragmentation.

\begin{figure}
\centering\includegraphics[scale=1]{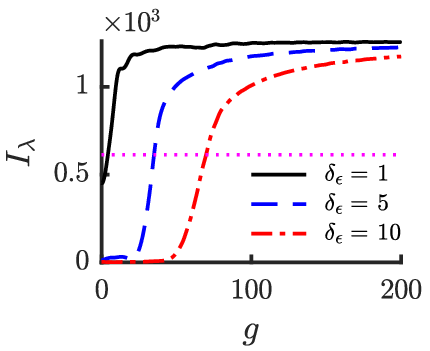}\hfill\includegraphics[scale=1]{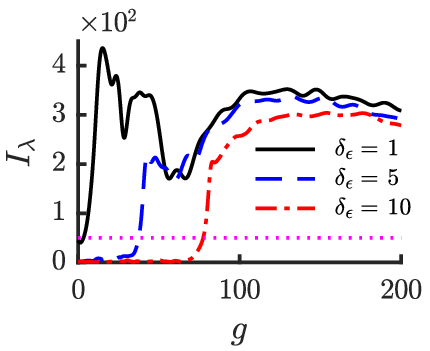}
\caption{QFI for a fragmented (left) and coherent (right) initial state as a function of the interactions strength ($\lambda=1$, $t=1$). The magenta dotted line corresponds to the QFI for a multiplicative Hamiltonian $\lambda \hj_x$ (ideal protocol regarding the maximal QFI), and 
helps to visualize how the natural dynamics of the system improves the estimation of the acceleration.  }
\label{fig:QFI_int}
\end{figure}

The full protocol would thus be represented by the following sequence:
\begin{compactenum}[a)]\itemsep0.25em
\item Prepare the system in its ground state.
\item Turn off the interaction during a time $\pi/(2\de)$ (beam splitting).
\item Apply the acceleration and tune the interaction to the value $g$, during a time $t$.
\item Apply the beam splitter again and perform the measurement.
\end{compactenum}
As we use the QFI, the last step is not relevant in our analysis. Formally we are interested in the calculation of the QFI of the state $\e{-\ii t (\hat{H}_\mm{sys}+\hat{H}_\mm{acc}(\lambda))}\e{-\ii \frac{\pi}{2} \hj_z} \ket{ \psi_\mm{g}(\theta)} $. The QFI can be calculated using the dynamical generator: 
\begin{equation}
\qfinue_\lambda  = 4(\moyvec{\hat{\msc{H}}^2}{\tilde{\psi}_\mm{g}(\theta)}-\moyvec{\hat{\msc{H}}}{\tilde{\psi}_\mm{g}(\theta)}^2 )\,,
\end{equation}
with $\ket{\tilde{\psi}_\mm{g}(\theta)}=\e{-\ii \frac{\pi}{2} \hj_z}\ket{\psi_\mm{g}(\theta)}$.

We represented in the figure \ref{fig:QFI_int} the QFI for both initial states as a function of the interaction strength. Regarding the fragmented state (left plot) the global behavior is similar to the behavior of the cQFI (compare plot on the top left in figure \ref{fig:cQFI_various}). 
Still we see that the maximum QFI achieved here is only half of the maximal cQFI. Moreover the QFI reaches high values only for relatively large interaction strength: For $\de=10$ we need  $g\gtrsim 100$ to reach the same order of magnitude as the maximal cQFI. Notice that at $g=200$ we are still in the expected regime of validity of the two-mode approximation, see section \ref{H}. 
In this plot, we also represented the QFI for an ideal interferometer (dotted magenta line). We see that owing to the presence of interaction in the dynamics the QFI can reach values roughly two times higher than the one obtained for an ideal interferometer with the same state.

On the right-hand plot we represented the QFI for the coherent state. There the behavior is much more erratic, especially for $\de=1$ and $\de=5$. For $\de=10$, we retrieve a behavior similar to what we already observed, meaning that at a given threshold the QFI increases steeply and then reaches what resembles a plateau. Importantly, we see that at its maximum the QFI is still one order of magnitude lower than the one obtained for the fragmented state. However, if we compare with the QFI obtained in an ideal interferometer  (magenta dotted line), 
we see that the relative gain is high, reaching a factor six for values of 
$g\gtrsim 100$.

%%%%%%%%%%%%%%%%%%%%%%%%%%%%%%%%%%%%%%%%%%%%%%%%%%%
%%%%%%%%%%%%%%%%%%%%%%%%%%%%%%%%%%%%%%%%%%%%%%%%%%%
\section{Conclusion}\label{sec:ccl}
%%%%%%%%%%%%%%%%%%%%%%%%%%%%%%%%%%%%%%%%%%%%%%%%%%%
%%%%%%%%%%%%%%%%%%%%%%%%%%%%%%%%%%%%%%%%%%%%%%%%%%%

We have investigated the possibility of estimating an external force applied to {\em scalar} bosons in a {\em single} potential well. 
Our analysis, using the model assumption of a two-mode truncation which is expected to be valid for weak to intermediate interaction strength, suggests a novel way to use cold atoms to perform quantum-enhanced metrology.  For scalar bosons in a single trap the splitting process into two modes is solely due to the interaction between the elementary constituents of the scalar Bose gas, and not due to external splitting (double well) nor due to internal splitting (spinor gas).

In terms of metrology, the crucial difference to other two-mode models lies in the dynamics. {In particular, the presence of interactions in our system leads in the Hamiltonian to both a renormalization of the single-particle term as well as to a nonlinearity whose detailed form depends on the mode functions.} Therefore, we have started by using the QFI maximized over initial state (cQFI) to characterize the system. We showed analytically that in the absence of interaction and in the regime where the two-mode truncation is expected to work (small accelerations), the cQFI --- and therefore the QFI based on any initial state --- is very far from its optimal value. We then verified numerically that the presence of interaction helps to increase the QFI to a large degree. This demonstrates,  independently from the initial many-body state, the necessity of finite interactions to benefit 
from a substantial quantum enhancement within our setup.

In a second part, we focused on realistic quantum many-body ground states. Namely we considered a fragmented state [a superposition of two spin-coherent states in SU(2) language] and a coherent state, two states that can in principle be obtained as ground states of our system. For both states, increasing the QFI requires a larger value of interaction strength, which is however still weak enough such that the two-mode truncation is expected to approximate the ground state. With the coherent state, the highest QFI observed is roughly equal to half the cQFI, while for the coherent state the ratio is closer to one tenth. 

The present results, which should be complemented by a fully self-consistent 
many-body approach, make a significant step towards exploring the possibility of 
quantum-enhanced estimation of an external force using scalar bosons in a single potential well.

\begin{acknowledgments}
We thank Seunglee Bae for his contributions at the early stages of this work and the anonymous 
Referee for useful suggestions.
This research was supported by the NRF of Korea, Grant No.~2017R1A2A2A05001422. 
\end{acknowledgments}

%\appendix

\bibliography{mbqm_v2_3.bib}

\end{document}